\documentstyle[floats,multicol,aps,epsf]{revtex}
\input epsf
\input psfig

   \newcommand{\be}{\begin{equation}}
   \newcommand{\ee}{\end{equation}}
   \newcommand{\bea}{\begin{eqnarray}}
   \newcommand{\eea}{\end{eqnarray}}

   \newcommand{\om}{\omega_n}
   \newcommand{\bd}{\begin{displaymath}}
   \newcommand{\ed}{\end{displaymath}}

\begin{document}
\draft
\widetext
%***********    This is for two columns *******************************
\twocolumn[\hsize\textwidth\columnwidth\hsize\csname @twocolumnfalse\endcsname
%********************************

\title{Non Magnetic Impurities in the Spin-Gap Phase of the Cuprates}
\author{Catherine P\'epin and Patrick A. Lee}
\address{Department of Physics, Massachussetts Institute of Technology, Cambridge, MA 02139}
\date{\today}

\maketitle \widetext
  \leftskip 54.8pt
  \rightskip 54.8pt
  \begin{abstract}
	It is now well established that Zn doping of high-$T_C$ cuprates
reduces their $T_C$ and triggers the appearence of a spin glass phase. In this
context, we have solved exactly the problem of N non magnetic impurities in
the staggered flux phase of the Heisenberg model which we assume to be a
good mean-field approximation for the spin-gap phase of the cuprates. In
this model, the quasiparticule spectrum has four nodes on the Fermi
surface, and linearization of the spectrum in the neighbourhood of these
nodes leads to a system of 2D Dirac fermions. In the presence of a
macroscopic number of (non magnetic) impurities, the problem has a
characteristic logarithmic structure that renders ineffective the usual
perturbative expansions. We have used this logarithmic structure to
calculate an exact solution. For a concentration ni of impurities in the
unitary scattering limit, the additional density of states is found to be
proportional to $ni/(w \ln^2 (|w|/D))$ (where D is the infrared cut-off of the
linearized spectrum) in analogy with the 1D case of doped spin-Peierls and
two-leg ladders compounds. We argue that the system exhibits a quasi
long-range order at $T=0$ with instantaneous spin-spin correlations
decreasing as $ni/ \ln^4 (ni/R_{ij})$ for large distances $R_{ij}$ between impurity
sites. We predict enhanced low energy fluctuations and compare these
results to NMR and inelastic neutron scattering experiments in the high-$T_C$
cuprates.
\par
  \end{abstract}
\vspace{0.2in}
]
%%%%%%%%%%%%%%%%%%%%%%%%%%%%%%%%%%%%%%%%%%%%%%%%%%%%%%%%%%%%%%%%%%%%%%%%%%%%%
\narrowtext

It is now established that the normal phase of the underdoped cuprates possesses a magnetic gap with the same anisotropy as the d-wave superconducting gap. The essential physics of the high-T$_C$ superconductors can be captured by focusing on the CuO$_2$ planes. A microscopic starting point of theoretical analysis is the so-called t-J model. In a slave-boson mean-field formulation, the system undergoes spin and charge separation: an electron in these highly correlated materials is a composite object made of a spin $1/2$ neutral fermion (spinon) and a spinless charged boson (holon). The gap in the magnetic excitations of the normal state can be viewed as singlet formation between pairs of neutral fermions in the absence of coherence between the holons. We assume that this mean-field picture captures essentially the initial idea of Anderson~\cite{anderson} of a Resonance Valence Bond ground state for the normal phase of the cuprates.
\par
Substitution of Cu ions in the conduction planes of high-T$_C$ cuprates by different non magnetic ions presents an important experimental tool for the study of the metallic state. Unusual effects has been revealed especially when the materials were doped with Zn. The valency of Zn is Zn$^{2+}$ (d$^{10}$) and compared with the Cu$^{2+}$ case, one electron is trapped by an additional positive charge of the nucleus, forming a singlet at the Zn site. In the spin-gap phase, it is experimentally found by NMR~\cite{xiao,alloul1,alloul2,mendels} experiments that a local magnetic moment of spin $1/2$ appears on Cu sites neighbouring the Zn impurity.
\par
In high-T$_C$ cuprates the persistance of AF fluctuations in the metallic state is probably one of the most striking feature. 
%In this light Inelastic Neutron Scattering experiments (INS) can be viewed as %a way to probe how closely related are magnetism and superconductivity in the %cuprates. 
Inelastic Neutron Scattering have established previously the existence of an energy gap in the imaginary part of the dynamical susceptibility in the normal and superconducting phases of the pure compound~\cite{neutrons} (without Zn). Substitution of Zn in CuO$_2$ planes shows a striking transfer of spectral weight from high to low energies, partially filling the spin-gap~\cite{41mev,kakurai}. This is the signature of strong enhancement of AF correlations in the spin-gap phase.
\par
Earlier work has treated the effect of a single non magnetic impurity in the $\pi$-flux state~\cite{autres1} and the pairwise interaction between them~\cite{autres2}. It was found that each impurity creates a bound state in the pseudogap at $\omega = 0$. Interaction between a pair of impurities leads to a level splitting between these states given by $\Delta(R) \sim1/\left ( R \ln (R) \right)$.

\par
Here we solve the problem of $N$ non magnetic impurities in the staggered flux phase of the Heisenberg model.
\par
In the mean-field of the $t-J$ model at half filling, the $\pi$-flux phase is a system of 2 dimensional Dirac Fermions with nodes located at the for points $\left ( \pm \pi/2, \pm \pi/2 \right )$. The impurities are treated as repulsive scalar potentials randomly distributed on the lattice.

\par
The key idea of this solution is to write the $T$-matrix equation directly for $N$ impurities and then to factorize the leading divergence. As the impurities are treated as scalar scattering centers, equations of motion can be closed algebraically by a $N \times N$ matrix ${\hat M}$ which caracterize the problem.

Here $M_{ij} \propto 4/ ( \pi D^2 ) i \om K_0 \left ( R_{ij} | \om | /  D \right )$ if $i$ and $j$ belong to the same sublattice and 

$M_{ij} \propto 4/( \pi D^2) i \om K_1 \left ( R_{ij} | \om | / D \right )$, if $i$ and $j$ belong to different sublattices. Now $ {\hat T}= -V_0 {\hat M}^{-1}$ and in the unitary limit, additional density of states can be casted into the form $\delta \rho (\omega) = -1/\pi Im Tr \left ( {\hat T} \partial {\hat M}/ \partial \omega \right )  = -V_0 / \pi Im Tr \left ( {\hat M}^{-2} \partial {\hat M }^2/\partial \omega \right )$.

All the configurations of impurities produce a logarithmic divergence in ${\hat M}^2$ that we factorize : ${\hat M}^2 \propto \ln |  \om R_{ij} / D | \ {\hat S} $, where ${\hat S}$ is a $N\times N$ matrix satisfying $S_{ij}=1 $ if $R_{ij} < D/|\om|$ and $0$ elsewhere.
We argue that the leading part of the density of states is due to the logarithmic divergence and the effect of ${\hat S}$ is negligeable. The value of ${\hat S}$ above is an approximation by excess of the rest. 

We take ${\hat S}^{-1}_{ij} = (|\om|/D) \ U_{ij}$. For all $j$ inside a circle of radius $D/|\om|$ around the point $i$, $U_{ij}$ is a random configuration of $\pm 1$ so that $\sum_j U_{ij} \sim  D/|\om|$. In addition all the points $j$ situated in the external boundary of this circle have $U_{ij}= -1/\pi$. Elsewhere $U_{ij}=0$. The main difficulty in the inversion of ${\hat S}$ is that two circles of radius $D/|\om|$ centered around two points $i$ and $j$ very close to each other will overlap, leading to the same number of non zero coefficients in the lines $i$ and $j$ of ${\hat S}$. In order to differentiate the sums $\sum_k S_{ik} S_{ki}^{-1}$ and $\sum_k S_{ik} S_{kj}^{-1}$ we have thus used the external boundary of the circle to compensate its volume in ${\hat S}^{-1}$.

With this inverse we show that the effect of ${\hat S}$ is negligeable in the density of states. The result is exact in the limit of low frequencies and in the unitary limit. We find
\be
\delta \rho ( \omega ) = \frac{n_i}{\omega \ln^2 | \frac{\omega}{D} | + (\pi/2)^2} \ .
\ee
 For one impurity alone~\cite{autres1,autres2}, a $\delta$-like bound state is created at $\omega=0$. If the impurities were totally uncorrelated, we would find $N$ $\delta$-functions at $\omega=0$. Here we see clearly the overlap of impurity states which leads to a broadening of the $\delta$ functions. It is interesting to compare this density of states with the one found for one dimensional spin-gap systems such as spin-Peierls and two-leg ladders systems where ${\displaystyle \delta \rho(\omega) \sim 1/ \left(| \omega | \ln^3( |\omega|/D ) \right ) }$.

\par
 In order to adress the question of long-range order in this system, we compute the instantaneous sorrelation function $ \left \langle S_i^+ S_j^- \right \rangle$.  We find that at $T=0$, 
\be
\label{eq20}
\langle S_i^+ S_j^- \rangle \sim \frac{n_i}{\left ( \ln^2\left ( n_i R_{ij} \right ) + (\pi/2)^2 \right) }\ ,
\ee
for impurities in different sublattices.
 $N$ non magnetic impurities randomly distributed in a $\pi$-flux phase stabilize a quasi long range staggered order. Physically we understand what happens by introducing impurities one after another. For one impurity, a local moment with $S=1/2$ is created in the vicinity~\cite{autres2}. Two impurities interact via an effective Heisenberg exchange $H=J(R) {\bf S}_1 . {\bf S}_2$, with $J(R) \sim 1/(R \ln(R))$ if they are located on different sublattices. However two impurities located on the same sublattice don't interact. This commensurability effect with sublattices gives the intuition that no frustration will be introduced in the model when a macroscopic number of impurities interact with each other. This effect is the origin of the quasi long range order that we find.

The instantaneous correlations on different sublattices are proportional to the density of impurities because a pair of impurities interact via an effective exchange $J(R) \sim 1/(R \ln(R))$. We suggest that the logarithmic decay in~(\ref{eq20}) is due to quantum fluctuations at zero temperature even though the precise power in the logarithm may depend on our approximation.
\par
What would be observed if a Neutron Scattering experiment would be performed on this system?  We have computed here the structure factor ${\bar S} ({\bf Q}, \Omega) = 1/{\cal N} \sum_q \chi^{\prime \prime}({\bf Q}, \Omega)$ integrated around $(\pi, \pi)$.
Using the same tricks as before for the instantaneous correlation function, we show that at low frequencies
${\bar S}({\bf Q}, \Omega) \sim n_i / \left ( \Omega \ln^4 \left | \Omega / D \right | \right ) $.
Even though we don't have true long range order in the system (which would have led to ${\bar S}({\bf Q}, \Omega) = \delta (\Omega)$ ), Neutron Scattering experiments would see a divergence of the integrated staggered structure factor at low frequencies, signaling a strong enhancement of antiferromagnetic correlations.

The situation will change for experimental systems with finite oxygen doping. If we take as a theoretical starting point the d-RVB ground state, the effect of oxygen doping will be to move the nodes away from the four points $(\pm \pi/2,\pm \pi/2)$ in the canonical spectrum. Excitations from node to node will then slightly differ from the commensurated vectors $(\pi,\pi)$ or $(\pi,0), (0,\pi)$. We believe that instead of quasi-long range AF order, this mechanism may give rise to an incommensurate quasi-long range order. Furthermore, it is likely that the finite frequency response integrated over $q$-space will not be that different from that we find, which may explain the enhancement of AF correlations observed in INS.

	\vspace{-.5  cm} % remove this before submission

% \end{document}

%\begin{figure}
%\centerline{\psfig{file=impfig2.eps,height=5cm,width=6cm}}
%\vspace{0.3cm}
%\caption{Sketch of the quasiparticle spectrum in a $\pi$-flux phase. The spect%rum has been linearised around the four nodes.}
%\label{fig1}
%\end{figure}

\end{document}